\documentclass[aps,prl,twocolumn,showpacs,superscriptaddress,groupedaddress]{revtex4}  
\usepackage{dcolumn}   
\usepackage{bm}        
\usepackage{amssymb}   
\usepackage{amsmath}
\usepackage{graphicx,tensor, cleveref}  
\usepackage{dsfont}
\usepackage{cancel}
\usepackage{xcolor}
\newcommand{\rev}[1]{{\color{black}#1}}

\begin{document}
\title{Optimal Control in Pandemics}
 \author{Joseph Samuel}
 \affiliation{Raman Research Institute, Bangalore-560080, India}
 \affiliation{International Center for theoretical Sciences,\\ Tata Institute of Fundamental Research, Bangalore-560089, India}
 \author{Supurna Sinha}
 \affiliation{Raman Research Institute, Bangalore-560080, India}



\date{\today}

\begin{abstract}
During a pandemic, there are conflicting demands arising from
public health and \textcolor{black}{socio-economic cost}. 
Lockdowns are a common way of containing infections,
but they adversely affect the economy. We study the question of how to 
minimise the \rev{socio-economic} damage of a lockdown while still containing
infections. Our analysis is based on the SIR model, which we analyse 
using a clock set by the virus. This use of the ``virus time'' 
permits a clean mathematical formulation
of our problem. We optimise the \rev{socio-economic cost} for a fixed
health cost  and arrive at a strategy for navigating the pandemic.
This involves adjusting the level of lockdowns in a controlled manner
so as to minimise the \rev{socio-economic} cost.

\end{abstract}

\maketitle



{\it Introduction}
The Covid-19 virus presents a global threat to life and livelihoods and 
throws up challenges which societies across the world have to learn to deal with. 
Pandemics are not new and there are mathematical models which have been developed over the years.
The value of mathematical models is that they give us a simplified picture of the pandemic and let us explore
the effects of different containment strategies, 
without performing costly and possibly fatal, social experiments. These models are the basis for a rational, science 
based social response to a serious threat. While models do have their limitations, they are steadily improving with time, experience and computational power 
\footnote{It is worth emphasizing that models for weather prediction which were unreliable
a few decades ago have now come of age and give us reliable predictions of the path  taken by a cyclone.}.
It is imperative for us to understand the predictions of these models and compare them with data and experience.

In this paper we consider one of the simplest models of disease spread, the SIR model\cite{ogilvymckendrickwalker}.
Our focus here is to quantify the social cost of a pandemic within the framework of the 
SIR model. As a society we would like to use interventions in order to minimise the damage caused by the disease.

By far the most important interventions are medical: doctors, nurses, medical infrastructure, treatments (drugs), 
preventive measures (vaccines), testing and contact tracing.
When a pandemic breaks out it takes time to develop some of these interventions.  Safe vaccines and drugs take time to test and develop. 
If the infections get out of hand, contact tracing too becomes impractical.  
We are then left with the non-medical interventions, like lockdowns,
which limit the spread of infection by changing the social behaviour of the population. This is the focus of the present paper.
Lockdowns limit the spread of disease by
reducing social contact; however, they also prevent the economy from functioning normally and so, come with a
\rev{socio-economic (SE) cost}. Like the health cost of a pandemic, the economic 
cost of a lockdown can be debilitating: lockdowns
affect lives and livelihoods, cause physical and mental trauma and even deaths.

Extreme strategies are
\begin{itemize}
\item to ignore the \rev{SE cost} and impose strict lockdowns (to the grievous detriment of the economy)
and
\item to ignore the health cost and keep the economy running normally (which results in a large human cost of suffering and death).
\end{itemize}
The \rev{SE cost} and the health cost are like Scylla and Charybdis of Greek mythology. We would
like to have a rational strategy of steering a course between these hazards, optimising the extent and timing
of lockdowns to minimise the total cost to society. In order to do this, we need
to model these costs in mathematical terms. Before we do that we recall the SIR model for disease spread. 

The SIR model divides the population 
into three compartments 
$\{S,I,R\}$, where $\{S,I,R\}$ are respectively the fractions
of susceptible, infected and removed populations. 
The removed population includes recoveries as well as deaths.
The model  assumes that the
recovered population is immune to the disease; that there is no possibility of reinfection.The progress of the disease is described by a set of three ordinary differential equations:

\begin{eqnarray}
\nonumber
\frac{dS}{dt}&=& -\beta(t) I S\\ \nonumber
\frac{dI}{dt}&=& \beta(t) I S -\gamma I\\ 
\frac{dR}{dt}&=& \gamma I, 
\label{sir}
\end{eqnarray}
where we allow for the possibility that 
$\beta$ varies with time, as would 
happen when lockdowns are imposed and relaxed. 
Evidently
\begin{equation}
S+I+R=1
\label{constraint}
\end{equation}
and the equations of the SIR model (\ref{sir}) maintain this condition. The progress of the disease can be described by a point in a two dimensional space,
a plane (\ref{constraint}) in the three dimensional $\{S,I,R\}$ space. 
 
The parameter $\beta$ describes the rate at which the 
Susceptible population becomes Infected due to contact with the Infected population.
This parameter 
depends on how infectious the disease is, as well as, the degree 
of contact between people. The effective infectivity $\beta(t)$ is a product  $\beta(t)=u(t) \beta_0$ \cite{dhar2020critique} of the biological infectivity $\beta_0$ 
(which depends on the disease) and $u(t)$, the degree of social contact between people. 
$\beta(t)$ can be controlled by 
reducing social contact $u(t)$, 
for example by using lockdowns to ensure physical distancing and
using masks.
$\gamma$ is the rate at which infected individuals either recover or die 
from the infection. Early detection and good 
medical care can increase the recovery rate.
$\beta$ and $\gamma$ which appear in the equations (\ref{sir}) are parameters
of the model which are both positive. $\gamma$ 
is assumed to be constant in time. The model is characterised essentially
by one parameter, the reproduction ratio $r=\frac{\beta}{\gamma}$. 
The independent time variable $t$ can be rescaled to set $\gamma$ to $1$.

The SIR model describes the evolution of the disease in a fixed population and is one of the  
simplest models, which captures the essential features of disease spread. More detailed compartmental models have also been studied. Among these
are the SEIR model and its variants \cite{abhicovid,science}, which have more compartments to allow for asymptomatic infections etc. 
There is also a study \cite{suvrat}, which questions the effectiveness of lockdowns in preventing fatalities. A suggestion for mitigating the
\rev{SE cost} of lockdowns has been made in Ref.\cite{israel}.
The independent variable in the SIR model is the time $t$ measured, say in days and there are three dependent variables $\{S,I,R\}$
subject to a single constraint (\ref{constraint}). 
Figure 1 shows the evolution of the SIR fractions as a function of time.
\begin{figure}[h!]
                \begin{center}
                        \includegraphics[width=0.4\textwidth]{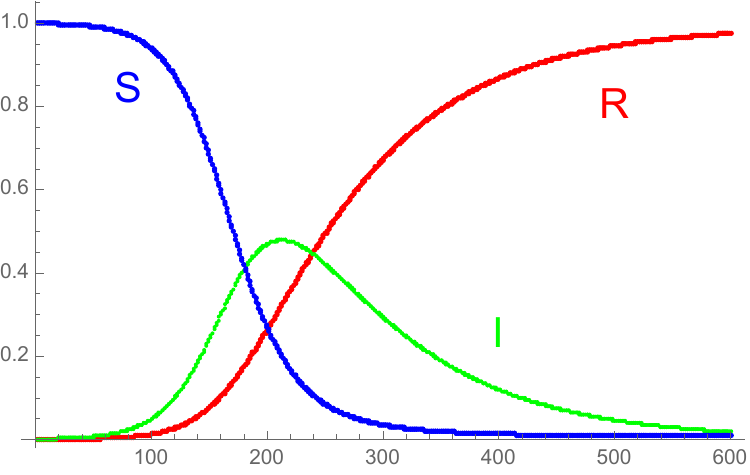}
                        \caption{The figure shows the 
susceptible (decreasing, blue online), 
infected (non-monotonic, green online) and removed (increasing, red online) 
fractions as a function of time 
in days. In this graph, for illustration we have taken $\beta=0.5$ and 
$\gamma=0.1$. This corresponds to a reproductive ratio of $r=5$.}
                \end{center}
        \end{figure}

This paper is organised as follows. We first summarise our main results and  
then present a derivation of our main results and finally we end with some concluding remarks.

{\it Main Results}
Here we summarise the main results of the study and describe the 
methods we use. Our objective is to minimise the damage caused by the pandemic
on two fronts: from the public health perspective and the economy. The 
demands of public health force us to impose lockdowns, which adversely affect
the economy. The question of interest is: when and how much to lock down
so that the damage to the economy is minimised. The question is complicated
by the fact that lockdown measures taken at a certain time 
can influence the infection
rates at later times. To understand this influence requires the use of a model
for the spread of infections. We work with the simplest SIR model.
In order to gain a long-term perspective, we have to 
consider the entire duration of the pandemic and account for the integrated 
health and \rev{SE costs}. 

This is precisely the kind of problem which can be dealt with 
using the calculus of variations. To give a familiar example, the shape
of a soap bubble is determined by the requirement that its surface area
is a minimum, subject to the constraint that the volume of enclosed air 
is fixed.
The shape which achieves this optimisation is the sphere. 
In the case of the pandemic, the role of the ``shape'' is played
by the profile of lockdown characterised by $\beta$ as a function of time, which tells 
when and how much to lock down. The role of the ``area'' is played by
the total integrated \rev{SE cost} of the lockdown. 
The role of the fixed volume of the soap bubble is played by the
total health cost, measured by the fraction of people affected over
the duration of the pandemic. What emerges from this study is
the optimal profile for $\beta(t)$; i.e that 
 which minimises the \rev{SE cost} for a fixed health cost. 
This is the analogue of the spherical shape of  a soap bubble.

A crucial ingredient in our study is the use of a 
new time variable. The time variable in the original SIR equations
(\ref{sir}) is human time. Human time is counted in  days or
weeks and measured by the progress of stars in the sky. In constrast,
virus time $\tau=R$ is measured by the progress of 
the virus through the population. The virus clock starts ticking
at the beginning of the pandemic, ($\tau=0$, when $R=0$),
runs faster when there are more infections ($\frac{d\tau}{dt}=\gamma I$)
and ceases to tick when the infections die out at the end of the pandemic.

The results of our study are presented in Fig.2,
which shows the optimal lockdown as a function of human time.
\begin{figure}[h!]
                \begin{center}
                        \includegraphics[width=0.4\textwidth]{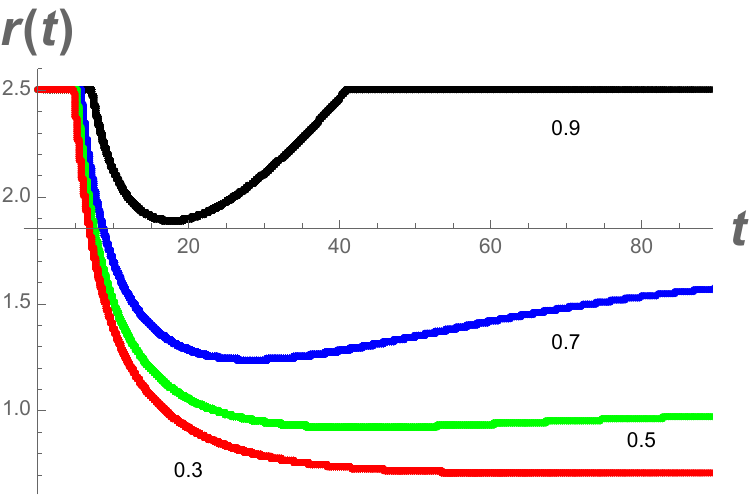}
                        \caption{Optimal lockdown profiles:
Here we plot the reproduction ratio $r(t)=\beta(t)/\gamma$ as a 
function of human time. The different
curves correspond to $\tau_f=0.3,0.5,0.7,0.9$ from bottom to top
corresponding to the colors (online) red, green, blue, black, respectively. $\tilde{\beta}$ and $\tau_0$ (see main text) have been set to .25 and 0.0 respectively 
and $\gamma$ to .1.}

                \end{center}
        \end{figure}
These curves are the main results of this study. They represent the optimal
way to modulate the lockdown so that the impact on the economy is minimal.
Each of these curves represents a different fixed health cost in terms of
the number of people affected by the virus.

{\it Derivation Of Main Results:}
Here we derive the main results of this paper. 
We express the health and \rev{SE cost} in mathematical form. 
\noindent
{\it \rev{Socio-Economic Cost}:}
The parameter in the SIR model 
which represents the effect of lockdown is $\beta$. We suppose that when 
all measures which do not affect the economy (like wearing masks, washing hands) 
have been imposed, we have 
$\beta= \tilde{\beta}$. 
Further reduction in $\beta$ can only come at an \rev{SE cost}. 
The \rev{SE cost} is a function that
decreases with increasing $\beta$ till $\tilde{\beta}$ and then 
drops down to zero. 
We make a simple choice of this function: 
for $\beta$ values less than $\tilde{\beta}$, the \rev{SE cost} of a lockdown 
is inversely proportional to $\beta$ and directly proportional 
to the number of days it lasts. We emphasise that we are not interested in a detailed modelling of the economy. We only wish to describe
the {\it damage} caused to the economy by the intensity and duration of the lockdown.

The total \rev{SE cost} integrated over the duration of the pandemic is 
modelled as \footnote{We could write this equation more correctly by multiplying the integrand by the Heaviside Theta function 
$\Theta(\tilde{\beta}-\beta)$. We have not done so as this may not be familiar to some readers.}
\begin{equation}
{\cal C}_E=\int_0^{\infty} \frac{dt}{\beta(t)}
\label{ecost}
\end{equation}
Note that the cost depends on the extent of the lockdown as well as the duration measured in human time $t$. 
Values of $\beta$ above $\tilde{\beta}$ can be ignored as they come with no \rev{SE cost}. If controlling the pandemic does not
require $\beta$ less than $\tilde{\beta}$, there is no conflict between the economic and public health objectives: the economy can function 
normally.
Below we assume that we are always dealing with $\beta$ values
less than $\tilde{\beta}$ i.e, there {\it is} a conflict between the twin objectives.

{\it Health Cost:} We model the health cost as $R_f=R(\infty)$ the total fraction of people affected by the disease during the entire course
of the epidemic. 
Hospitalisations and deaths are some fixed fractions of $R_f$. Even some of those who do not need hospitalisation suffer
long term after effects from the ravages of Covid-19. We can therefore model the health cost mathematically as
$R_f$, the final value of the Removed fraction:
${\cal C}_H= R_f$. ${\cal C}_H$ is a  dimensionless number. 
Our objective is to hold ${\cal C}_H$ fixed at the value ${\cal C}_{H0}$ and find the lockdown profile $\beta(t)$ which minimises the 
\rev{SE cost}. 
The fixed value ${\cal C}_{H0}$ of the health cost is a choice one has to make.
Needless to say, there is a value judgement involved
in making this choice. Choosing a small value for ${\cal C}_{H0}$ gives more weightage to the health cost 
and a large value reverses the emphasis.
Once this value judgement is made, we can use our ability to modulate $\beta$ over time, 
varying the extent and timing of lockdowns to minimise the 
\rev{SE cost}.

The independent variable $t$ in the SIR equations (\ref{sir}) is the time
measured in human time, for instance, days.
This is relevant to the progress of the
epidemic in human terms. In fact, the \rev{SE cost} (\ref{ecost})
grows with the duration of a lockdown, measured in human time. 
We find it advantageous to use a new time variable
as set by the progress of the virus through the human population.
Accordingly we set $\tau=R$ the fraction of
people affected by the epidemic and regard 
this to be the ``virus time''. We will use $\tau$ and $R$ interchangeably,
preferring $\tau$, when we wish to emphasise its role as a ``time'' or independent variable.
The virus time increases monotonically
\begin{equation}
\frac{d\tau}{dt}=\frac{d R}{dt}=\gamma I\ge0
\label{jacobian}
\end{equation}
with human time, the rate of progress 
given by $\gamma I>0$, which is proportional to the current infected fraction.
As we will see, using the virus time instead of the human time 
gives us significant advantages in addressing our problem. 
First, it gives us an exact parametric solution of the
SIR model. (This is equivalent to the parametric 
solutions given earlier by \cite{harko,miller}).  Second, we get
a clean mathematical formulation of our problem of optimising
the total social cost. 
Letting the virus set the clock is one of the
crucial ingredients of our approach.

Dividing the first of the equations Eq. (\ref{sir}) by the last,
we find that (expressing $\beta$ as a function of $\tau$)
\begin{equation}
\frac{dS}{d\tau}=\frac{dS}{dR}=-\frac{\beta(\tau)}{\gamma} S
\end{equation}
which is readily integrated. Using the constraint (\ref{constraint})
immediately gives us an exact solution of the SIR model in parametric form.
\begin{eqnarray}
\nonumber
S(\tau)&=& S_0 \exp{[-\frac{\int_{\tau_0}^{\tau}\beta(\tau') d\tau'}{\gamma}]} \\ \nonumber
I(\tau)&=& 1-S_0 \exp{[-\frac{\int_{\tau_0}^{\tau}\beta(\tau') d\tau'}{\gamma}}]-\tau\\
R(\tau)&=& \tau,
\label{sirsol}
\end{eqnarray}
where the last equation is a tautology arising from the definition of $\tau$.
The virus time $\tau$ is measured from the beginning of the epidemic. $\tau_0$ represents any fixed intermediate time. The relation between the virus
time and human time is given by integrating the last of (\ref{sir}), where
$I(\tau)$ is given by the second equation of (\ref{sirsol}).
\begin{equation}
t=\int_0^{\tau} \frac{d \tau'}{\gamma I(\tau')}
\label{tautot}
\end{equation}

Let us suppose that our lockdown response starts 
when the virus time is $\tau_0$,
when the Suceptible fraction is $S_0$. $\tau_0$ could be  
when the pandemic is initially detected or any subsequent time.
Given a fixed value of the health cost $R_f$, 
our problem is to choose 
the function $\beta(\tau)$ so as to minimise the \rev{SE cost}. 

Let us introduce a new variable
\begin{equation}
y(\tau)=\int_{\tau_0}^\tau \beta(\tau') d\tau'
\label{ydef}
\end{equation}
so that $S(\tau)=S_0\exp{[-y(\tau)/\gamma] }$ and 
$\beta(\tau)=\frac{dy}{d\tau}=\dot{y}$. 
Then $S_f=S(\tau_f)=S_0\exp{[-y(\tau_f)/\gamma}]$ and at the 
end of the pandemic, when infections cease
($I=0$), we have from $S_f+R_f=1$, 
\begin{equation}
S_0\exp{[-y(\tau_f)/\gamma}]+\tau_f=1
\label{fixing}
\end{equation}
fixing $y_f$ in terms of $\tau_f$.
\begin{equation}
y_f=y(\tau_f)=-\gamma{\log{(1-\tau_f)/S_0}}.
\label{fixing2}
\end{equation}

We now have a classic variational problem for $y(\tau)$, where $y(\tau_0)=0$, 
$y(\tau_f)=y_f$ are held fixed and we have to minimise
the \rev{SE cost}
\begin{equation}
{\cal C}_E= \int_{\tau_0}^{\tau_f} \frac{dt} {d\tau} \frac{d\tau}{\dot{y}}
\label{ecost2}
\end{equation}
From the parametric solution to the SIR equations (\ref{sir}), 
we replace $\frac{dt}{d\tau}$ by
$(\gamma I(\tau))^{-1}$
leading to the variational problem of minimising 
\begin{equation}
\int_{\tau_0}^{\tau_f}\frac{d\tau}{I[y(\tau)]\dot{y}(\tau)}
\label{eulerlag}
\end{equation}
where we have dropped some constants. $I[y(\tau)]$ here is a functional of $y(\tau)$ which is found by solving the SIR equations
(\ref{sir}). Its explicit form is given by 
the second of (\ref{sirsol}).
We now have to minimise
\begin{equation}
\int_{\tau_0}^{\tau_f}\frac{d\tau}{\dot{y}(\tau)(1-S_0\exp{[-y/\gamma]} -\tau)}
\label{eulerlag2}
\end{equation}
We now vary $y(\tau)$ in (\ref{eulerlag2}) and as is usual in 
the calculus of variations, integrate by parts and discard the boundary
terms, since $y$ is held fixed at both boundaries. 

We can read off the Lagrangian appearing in (\ref{eulerlag2}):
\begin{equation}
L(y,\dot{y},\tau)=\frac{1}{\dot{y} (1-S_0\exp{[-y/\gamma]} -\tau)}.
\label{lagrange}
\end{equation}
and the resulting Euler-Lagrange equations can be rearranged to give
\begin{equation}
{\ddot{y}}(1-S_0\exp{[-y]/\gamma} -\tau) + \frac{{\dot{y}}^2 S_0\exp{[-y/\gamma]}}{\gamma} - \frac{\dot{y}}{2} = 0
\label{eulerlag3}
\end{equation}

This equation can be integrated by expressing it as
\begin{equation}
\frac{d K(y,\dot{y},\tau)}{d\tau}=0
\label{dkdtau}
\end{equation}
where the constant of the motion $K$ has the form
\begin{equation}
K(y,\dot{y},\tau)=\dot{y} (1-S_0\exp[-y/\gamma]-\tau)+y/2=K_0
\label{constant}
\end{equation}

These equations can be analytically solved by introducing 
an integrating factor $(y/2-K)^{-3}$. The solution gives $\tau$ as a function
of $y$ expressed in terms of elementary functions including 
the exponential integral,
which can be plotted to show $y$ as a function of $\tau$. From this it is easy to extract the quantity of interest: 
$\beta(\tau)=\dot{y}$, which determines the lockdown profile. 

In making Fig.2 , we have numerically integrated (\ref{constant})
in the form 
\begin{equation}
\frac{dy}{d\tau}=\frac{K_0-y/2}{(1-S_0\exp[-y/\gamma]-\tau)}
\label{numerical}
\end{equation}
and noted that our boundary conditions imply that the value of the constant
is $K_0=y_f/2$.

Fig. 2 shows the optimal way of imposing lockdowns, 
plotting $r=\beta/\gamma$ as a function of human time (Fig.4). 
The optimal solution consists of an initial sharp lockdown followed by
a gradual release of the lockdown. Intuitively, this is easy to understand: premature release of
lockdown results in flare-ups of the disease, which then require further lockdowns which contribute to the \rev{SE cost}.

{\it Concluding Remarks}
First, a disclaimer: the authors of this paper are not epidemiologists. 
We are theoretical physicists who have addressed a 
socially relevant interdisciplinary 
problem, which can be addressed using the methods of our subject. 
As is common in theoretical physics, we work with the simplest model that captures the phenomena of interest.
The conclusions we arrive at are not intended to be {\it directly} transferred to any real world context. 
Nevertheless, the ideas developed here can be developed further by introducing more realistic models.
The main message we have to offer is that there is a competition between the twin social objectives of public  
health and the economy. It is then advantageous to use lockdown profiles derived from our formalism 
to minimise the total damage from a pandemic. 
For example, the SEIR model is a slightly more realistic 
model in which our analysis can be carried out. 
In this case, a purely analytic solution is not possible, but one can formulate the problem as we have done here
and derive results for the optimal lockdown using numerical methods.

One of the main ideas introduced and used in this paper is that of ``virus time''. At one level, it is a convenient mathematical
device. It leads to an exact parametric solution of the SIR model. This solution is considerably simpler than the existing ones\cite{exact}.
At a conceptual level, it is a more appropriate measure of the progress of the disease than human time measured in days. 
For example, virus time elapses  differently in different countries. Some countries impose travel restrictions (eg. travel within 5 Km of one's house). In such cases
virus time elapses differently in different locations.

Our graphs show the optimal lockdown profile in terms of the
reproductive ratio $\beta/\gamma$, a quantity which is directly measurable (by testing a random sample of the population)
and often used to describe the progress of a pandemic. 
One could
consider more complicated functional dependence on $\beta$ for the 
\rev{SE cost}, for instance, one could consider
the \rev{SE cost} per day to be inversely proportional to $\beta^2$ and so on. We expect the main qualitative
conclusions to remain unchanged by such a choice. It would take economists to realistically measure
the cost of lockdowns. This is a task we do not undertake here.

Across the world there have been many disparate government 
responses to the covid-19 pandemic. Sweden chose not to lock down at all;
New Zealand chose to lock down hard and early. 
Many other nations adopted policies in between, some responding to 
flaring infections as they broke out, as firemen do with fires. 
After the pandemic is over, with the clarity of hindsight 
we will learn which of these 
strategies was most effective in preventing societal distress. 
Meanwhile we can gain insights by 
working with simple models to evaluate these different 
strategies. 
The work of this paper is a starting point in considering the 
\rev{SE} as well as health costs of a pandemic. 
We import methods from the variational calculus (which we illustrate 
by the example of soap bubbles) to arrive at an optimal strategy to navigate the pandemic keeping both the \rev{SE} and health costs in mind.
A crucial ingredient in our problem is the virus time which is a natural parameter which has been introduced to make the problem tractable.
Our main result is that the best strategy to follow is one in 
which a sharp lockdown is imposed which is followed by a gradual release. This is indeed to be expected, if one understands exponential growth of disease. However,
our solution also prescribes when to lock down and how much. 
An important choice to be made in determining the strategy is the value of $R_f$, the health cost. This choice will depend on the \rev{resources} of the nation. 
Nations which can afford a larger \rev{cost} can opt for a lower value of $R_f$. Poorer nations will be forced to accept a higher health cost. 
However, given these limitations, the strategy we propose is 
optimal within the SIR model. 

{\it Acknowledgements}
It is a pleasure to thank 
Abhishek Dhar, Suvrat Raju, Biman Nath and Michael Berry for 
their comments on the manuscript. \rev{We thank the referees 
for their critical remarks which have helped us improve the 
manuscript.} 
\bibliography{active}

\end{document}